\documentclass{article}

\usepackage{arxiv}

\usepackage[utf8]{inputenc} 
\usepackage[T1]{fontenc}    
\usepackage{hyperref}       
\usepackage{url}            
\usepackage{booktabs}       
\usepackage{amsfonts}       
\usepackage{nicefrac}       
\usepackage{microtype}      
\usepackage{graphicx}
\usepackage{amsmath}

\title{Uncovering differential identifiability in network properties of human brain functional connectomes}

\date{} 					

\author{
  Meenusree ~Rajapandian\\
  School of Industrial Engineering\\
  Purdue University\\
  West Lafayette, IN 47906 \\
   \And
  Enrico ~Amico \\
  School of Industrial Engineering\\
  Purdue Institute of Integrative Neuroscience \\
  Purdue University\\
  West Lafayette, IN 47906 \\
   \And
  Kausar ~Abbas \\
  School of Industrial Engineering\\
  Purdue Institute of Integrative Neuroscience \\
  Purdue University\\
  West Lafayette, IN 47906 \\
   \And
  Mario ~Ventresca \\
  School of Industrial Engineering\\
  Purdue University\\
  West Lafayette, IN 47906 \\
   \And
  Joaqu\'{i}n Go\~{n}i \thanks{Use footnote for providing further
    information about author (webpage, alternative
    address)---\emph{not} for acknowledging funding agencies.} \\
  School of Industrial Engineering \\
  Purdue Institute of Integrative Neuroscience \\
  Weldon School of Biomedical Engineering \\
  Purdue University\\
  West Lafayette, IN 47906 \\
  \texttt{jgonicor@purdue.edu} \\
}

\begin{document}
\maketitle

\begin{abstract}
The Identifiability Framework ($\mathbb{I}\mathit{f}$) has been shown to improve differential identifiability (reliability across-sessions and -sites, and differentiability across-subjects) of functional connectomes for a variety of fMRI tasks. But having a robust single session/subject functional connectome is just the starting point to subsequently assess network properties for characterizing properties of integration, segregation and communicability, among others. Naturally, one wonders if uncovering identifiability at the connectome level also uncovers identifiability on the derived network properties. This also raises the question of where to apply the $\mathbb{I}\mathit{f}$ framework: on the connectivity data or directly on each network measurement? Our work answers these questions by exploring the differential identifiability profiles of network measures when $\mathbb{I}\mathit{f}$ is applied on 1) the functional connectomes, and 2) directly on derived network measurements.

Results show that improving across-session reliability of FCs also improves reliability of derived network measures. We also find that, for specific network properties, application of $\mathbb{I}\mathit{f}$ directly on network properties is more effective. Finally, we discover that applying the framework, either way, increases task sensitivity of network properties. At a time when the neuroscientific community is focused on subject-level inferences, this framework is able to uncover FC fingerprints, which propagates to derived network properties.
\end{abstract}

\keywords{brain connectomics, functional connectivity, fingerprint, network science, subject identifiability}

\section{Introduction}
The analysis of structural and functional human brain connectivity based on network science has become prevalent for understanding the underlying mechanisms of the human brain. Using network properties, we are able to understand the topology of brain connectivity patterns \cite{sporns2010networks, fornito2016fundamentals, sporns2018graph}, integration and segregation \cite{sporns2016modular, cohen2016segregation, deco2015rethinking, fukushima2018structure, sporns2013network}, as well as communication dynamics \cite{costa2011communication, estrada2008communicability, petrella2011use, avena2018communication} and association between human cognition and brain function \cite{zalesky2010network, davison2015brain, bola2015dynamic, alavash2015persistency, mattar2016flexible}. Until recently, many brain connectivity studies used group-level comparisons, where data from many subjects are collapsed (e.g. group averaging) into a representative sample of clinical and healthy population \cite{fornito2015connectomics, castellanos2013clinical, crossley2014hubs}. However, this comes at a price of potentially ignoring intra-group individual variability \cite{seitzman2019trait}.

Detecting individual differences in functional connectivity profiles thus becomes important, when associating connectivity profiles with individual behavioral outcomes. In recent years, publicly available functional connectome datasets \cite{van2013wu, biswal2010toward} with large sample sizes have enabled the scientific community to account for inter-individual variability in the human functional connectome (FC). A number of promising methods that can successfully capture these individual differences have been established in recent times \cite{mars2018connectivity, satterthwaite2018personalized, gratton2018functional, seitzman2019trait, venkatesh2019comparing}. For instance, work by \cite{finn2015functional} has shown the existence of a recurrent and reproducible fingerprint in functional connectomes estimated from neuroimaging data. This idea has been extended to maximize or minimize subject-specific and/or task-specific information \cite{pallares2018extracting, xie2018whole}. These subject-specific fingerprints have been used to track fluctuations in attention at the individual level \cite{rosenberg2019functional}.

The ``Identifiability Framework" \cite{amico2018quest}, based on the group-level Principal Component Analysis of functional connectomes that maximizes differential identifiability, has been shown to improve functional connectome fingerprints within- and across-sites, for a variety of fMRI tasks, over a wide range of scanning length , and with and without global signal regression \cite{bari2019uncovering,amico2018quest}. Additionally, it has been shown that maximising differential identifiability on the functional connectomes provides more robust and reliable associations with cognition \cite{svaldi2019optimizing} as well as with disease progression \cite{svaldi2018towards}. The natural next step is to assess the impact of such a procedure on subsequent network measurements that characterize topological and communication properties of functional brain networks.

An open question of great relevance for the Brain Connectomics community is how to measure and uncover subject fingerprints in network measurements of functional connectivity. Uncovering reliable connectivity fingerprints is crucial when assessing clinical populations and when ultimately mapping cognitive characteristics into connectivity \cite{svaldi2018towards,scheinost2019ten, shen2017using}. Our hypothesis is that improvement in FC fingerprint should also ``propagate" to network derived measurements. An organic way of assessing this would be to track differential identifiability scores of derived network features as the differential identifiability on the functional connectomes changes. One could also proceed with the application of the Identifiability Framework directly on the network derived features as opposed to using it on FCs. The above mentioned approaches rely on different principles of what is a fingerprint in a network derived measurement. The first one assumes that functional connectivity data is ``holding" the fingerprints and propagating them to any network derived measurement. The second one considers functional connectivity data as a proxy to ultimately estimate a network measurement with a potentially prominent subject fingerprint.

\section{Methods}

The dataset used here is the 100 unrelated subjects of the Human Connectome Project Release Q3 \cite{van2013wu}. Per HCP protocol, all subjects gave written informed consent to the HCP
consortium. Each subject consists of two fMRI resting state runs and seven fMRI tasks - gambling, relational, social, working memory, motor, language and emotion. Data acquisition for each subject and for each task consists of two fMRI sessions, which are tagged here as test and retest. A cortical parcellation into 360 brain regions as proposed by \cite{glasser2013minimal} was employed with an additional 14 sub-cortical regions for completeness \cite{amico2018mapping, amico2018quest}. The HCP functional preprocessing pipeline was used \cite{glasser2013minimal, smith2013resting}, followed by further processing as described in \cite{amico2018quest,amico2019centralized} for both resting state and task fMRI data. For each subject and fMRI session, a symmetric weighted connectivity matrix (the functional connectome) was obtained by computing Pearson's correlation coefficients between pairs of nodal time courses. For a detailed description of all the preprocessing steps, refer to \cite{amico2018quest}.


\subsection{Network Properties}

Graph theoretic measures have played a key role in understanding the attributes of brain networks in general, and of functional connectomes in particular \cite{sporns2010networks, fornito2016fundamentals, rubinov2010complex}. Here we select a set of node and node pair properties (i.e. properties that are a function of a single node or a pair of nodes respectively) to assess their fingerprinting characteristics. A functional connectome is a symmetric square correlation matrix that may be seen as an undirected weighted graph. Let $G=(V,\mathcal{W})$ be an undirected weighted graph with set of nodes $V= \{v_1,v_2,\dots,v_n\}$ and weights $ \mathcal{W} = [w_{ij}]$ where $w_{ij}$ is the strength of the edge between nodes $v_i$ and $v_j$.
 
\begin{enumerate}
\item Degree Strength

The degree strength of a node ($K_i$) in an undirected binary graph is the number of edges that are connected to the node. Here, we consider the weighted sum of the edges connected to the node $i$.

$$K_i = \sum_{j=1}^n w_{ij} $$

\item Shortest Path Length

The shortest path length (SPL) between two nodes of an undirected graph is defined as the minimum number of edges (and thus steps) that separate the two nodes. For an undirected weighted graph it is the path that results in the smallest value of the sum of the weights of the edges that constitute a path between a pair of nodes $i$ and $j$. For such a path, that consists of the following sequence of nodes, $\Omega_{i\leftrightarrow j} = \{i,x,y,\dots,z,j\}$ with corresponding sequence of edge weights $\pi_{i\leftrightarrow j} = \{w_{ix},w_{xy},\dots,w_{zj}\}= $, the shortest path length is

$$SPL_{ij} = \sum_{w_{lm} \in \pi_{i\leftrightarrow j}} \frac{1}{w_{lm}}$$

Note that $\Omega_{i\leftrightarrow j} = \Omega_{j\leftrightarrow i}$ for shortest paths in any undirected graph.

\smallskip
\item Search Information

The search information ($SI_{ij}$) for two nodes $i$ and $j$ is the information required to follow the shortest path \cite{rosvall2005networks} i.e. the negative log of the product of probability of taking the correct exit at every node along the shortest path. In other words, it can be considered as the information required to reach node $j$ starting from node $i$. For a path between nodes $i$ and $j$ that has a sequence of nodes $\Omega_{i\rightarrow j} = \{i,x,y,\dots,z,j\}$, with probability of taking the path $P(\pi_{i \rightarrow j}) = \Pi_{l \in \Omega^*_{i\rightarrow j}} 1/k_l$, the search information for the path is \cite{goni2014resting}

$$ SI_{ij} = -\log_2{P(\pi_{i \rightarrow j})}$$

Note that $SI_{ij} \neq SI_{ji}$
\smallskip

\item Mean First Passage Time

The mean first passage time (MFPT) is the expected (on average) number of steps a random walker could take to reach node $j$ (for the the first time) from node $i$ \cite{kemeny1976markov}. The Mean First Passage Time (MFPT) for a pair of nodes with source $i$ and target $j$ is

$$ MFPT_{ij} = \frac{\zeta_{jj} - \zeta_{ij}}{\phi_{j}}$$

where $\phi$ is the left eigenvector associated with eigen value $1$, $Z =[\zeta_{ij}]$ is the fundamental matrix computed as $Z = (I-P+\Phi)^{-1}$. Here $I$ is the $n\times n$ identity matrix, $P$ is the transition matrix and $\Phi$ is an $n \times n$ matrix with each column corresponding to the probability vector $\phi$ such that $\forall j$ $\Phi_{ij} = \phi_i$. Please note that $MFPT_{ij}\neq MFPT_{ji}$.

\smallskip
\item Driftness

We use a measure of communication called driftness \cite{costa2011communication} which is the ratio of the mean first passage time and the shortest path of a pair of nodes $i$ and $j$. Considering that $SP_{ij}$ is the best possible scenario path for a random-walk, this measurement is modulating the mean first passage times with respect to the fastest routes within the network to go from node $i$ to $j$. Hence, note that $W_{ij} \geq 1$.

$$W_{ij} = \frac{MFPT_{ij}}{SP_{ij}}$$

\item Communicability

Communicability between two nodes $i$ and $j$ is a measure of network integration computed as a weighted sum of number of all possible walks between them. \cite{estrada2008communicability} Here, we use a normalization method proposed to handle the disproportionate influence of highly connected nodes (also known as hubs) in a graph \cite{crofts2009weighted}. Note that this is frequently the case when assessing functional connectomes.

$$C_{ij} = [e^{D^{-0.5} A D^{-0.5}}]_{ij}$$

where $D=diag(K)$ and $K = [k_i]$ where $k_i$ is the degree strength of node $i$, as defined above.
\smallskip
\item Clustering Coefficient

The clustering coefficient of a node is the tendency of its neighbors to form cliques. It is the ratio of the total number of triangles a node forms with its neighbors to the total number of possible triangles that can be formed.

$$CC_i = \frac{2t_i}{k_i(k_i - 1)}$$

where $t_i = \frac{1}{2}\sum_{j,h \in V}(w_{ij}w_{ih}w_{jh})^{1/3}$ is the geometric mean of triangles around node $i$ for weighted networks.
\smallskip
\item Betweeness Centrality

The betweenness centrality of a node is the fraction of all shortest paths in a network that contain that node.

$$B_i = \frac{1}{(n-1)(n-2)}\sum_{\substack{h,j \in V \\ h\neq j, h\neq i, j\neq i}} \frac{\rho_{hj}(i)}{\rho_{hj}}$$

where $\rho_{hj}(i)$ is the number of shortest paths between $h$ and $j$ that pass through $i$.
It can be seen as a measurement of to what extent a node ``lies" between other pairs of nodes when accounting specifically for shortest-paths.

\end{enumerate}

\subsection{Group-level Principal Component Analysis and Differential Identifiability}

Briefly describing the Identifiability Framework ($\mathbb{I}\mathit{f}$) introduced in \cite{amico2018quest}, the functional connectomes of each subject (test and retest) are vectorized and added to a matrix, the columns of which are the runs (test and retest) of each subject, while the rows are the functional connectivity values of brain region pairs. The $m$ principal components of this matrix are then ranked by variance explained and included, in an iterative fashion, to reconstruct the functional connectomes \cite{amico2018quest}. This is done separately for each task and rest. Following the reconstruction of the functional connectomes, we then compute the network property of interest for each subject, on each run (test and retest). This is referred to as $NP(\mathbb{I}\mathit{f}\{FC\})$ in all further sections where $NP$ is the network property and $FC$ is the functional connectome.


We also extend the framework by using this decomposition - reconstruction procedure on the network properties. In this case, the network properties are computed on the original functional connectomes for each subject and run. Each network property is then vectorized and added to a matrix. Note that this is similar to how functional connectomes were rearranged in the $NP(\mathbb{I}\mathit{f}\{FC\})$ and in \cite{amico2018quest}. However, the rows of this matrix now consists of the network property values corresponding to a pair of brain regions in case of pairwise properties or a brain region when node properties are derived. The principal components of this matrix are then extracted and iteratively reconstructed using $m$ number of components with highest explained variance. Since the network properties are the ones being decomposed in this case, the result of the reconstruction are the corresponding network properties of each individual and each run. This method is subsequently referred to as $\mathbb{I}\mathit{f}\{NP(FC)\})$.

We use differential identifiability \cite{amico2018quest} to asses the individual fingerprint of each network property. For each method described above, the network properties derived are used to compute the identifiability matrix. Each position of the identifiability matrix $i,j$ denotes the correlation between the network property of subject $i$ test and subject $j$ retest. Then, along the diagonal elements, we have the correlation of a network property between the subject test and retest called $\mathit{I_{self}}$. The non-diagonal elements are the correlations between a run of a subject $i$ and subject $j$ where $i$ and $j$ are different ($\mathit{I_{others}}$). The differential identifiability is then defined as,

\begin{align*}
    \mathit{I_{diff}} = (\mathit{I_{self}} - \mathit{I_{others}}) * 100
\end{align*}


Intraclass correlation coefficient (ICC) represents how strongly measures of a group are in agreement with each other \cite{mcgraw1996forming, bartko1966intraclass}. The higher the ICC value, higher is the level of agreement. We use ICC \cite{shrout1979intraclass} to asses the task sensitivity of a network measure, for each brain region pair and every subject. In this case, the members of the groups are the different runs (test and retest) of a subject; the different groups represent the different fMRI task conditions (and rest). The mean task sensitivity is then taken across all subjects and reported. For this assessment, the functional connectome (or the network property $\mathbb{I}\mathit{f}\{NP(FC)\}$) was optimally reconstructed, i.e. using the number of components that gave the highest $\mathit{I_{diff}}$ score for that task.

\section{Results}
The dataset used for this study consisted of fMRI scans of the 100 unrelated subjects from the Human Connectome Project \cite{van2013wu}. For each subject, we computed  $18$ whole-brain functional connectivity matrices: $4$ corresponding to resting-state (2 sessions, each with test and retest), and $14$ corresponding to each of the $7$ tasks (each including two runs;  test-retest). The multimodal parcellation used here, as proposed by \cite{glasser2016multi}, includes $360$ cortical brain regions. For completeness, $14$ subcortical regions were added \cite{amico2018mapping}, hence producing functional connectome matrices (square, symmetric) of size $374 \times 374$.


In this work, we study the effects of $\mathbb{I}\mathit{f}$ on the identifiability profiles of network properties in two different scenarios: 1) when applying differential identifiability on functional connectivity, $NP(\mathbb{I}\mathit{f}\{FC\})$ and 2) when applying differential identifiability directly on network properties, $\mathbb{I}\mathit{f}\{NP(FC)\}$.  

$NP(\mathbb{I}\mathit{f}\{FC\})$: The functional connectomes (FCs) of each task (including rest) were vectorized, organized together and then decomposed into principal components and subsequently reconstructed by adding an increasing number of components ordered by their variance explained. After every such reconstruction, a number of network measurements [see Methods for details] were computed for each FC and $\mathit{I_{diff}}$ was found on the derived network properties. This is compared with the $\mathit{I_{diff}}$ score estimated directly from the reconstructed functional connectomes - $\mathbb{I}\mathit{f}\{FC\}$. By doing so, we extend the differential identifiability framework to uncover fingerprints in network properties derived from functional connectomes.

\begin{figure}[h!]
\centering
\includegraphics[width=\textwidth]{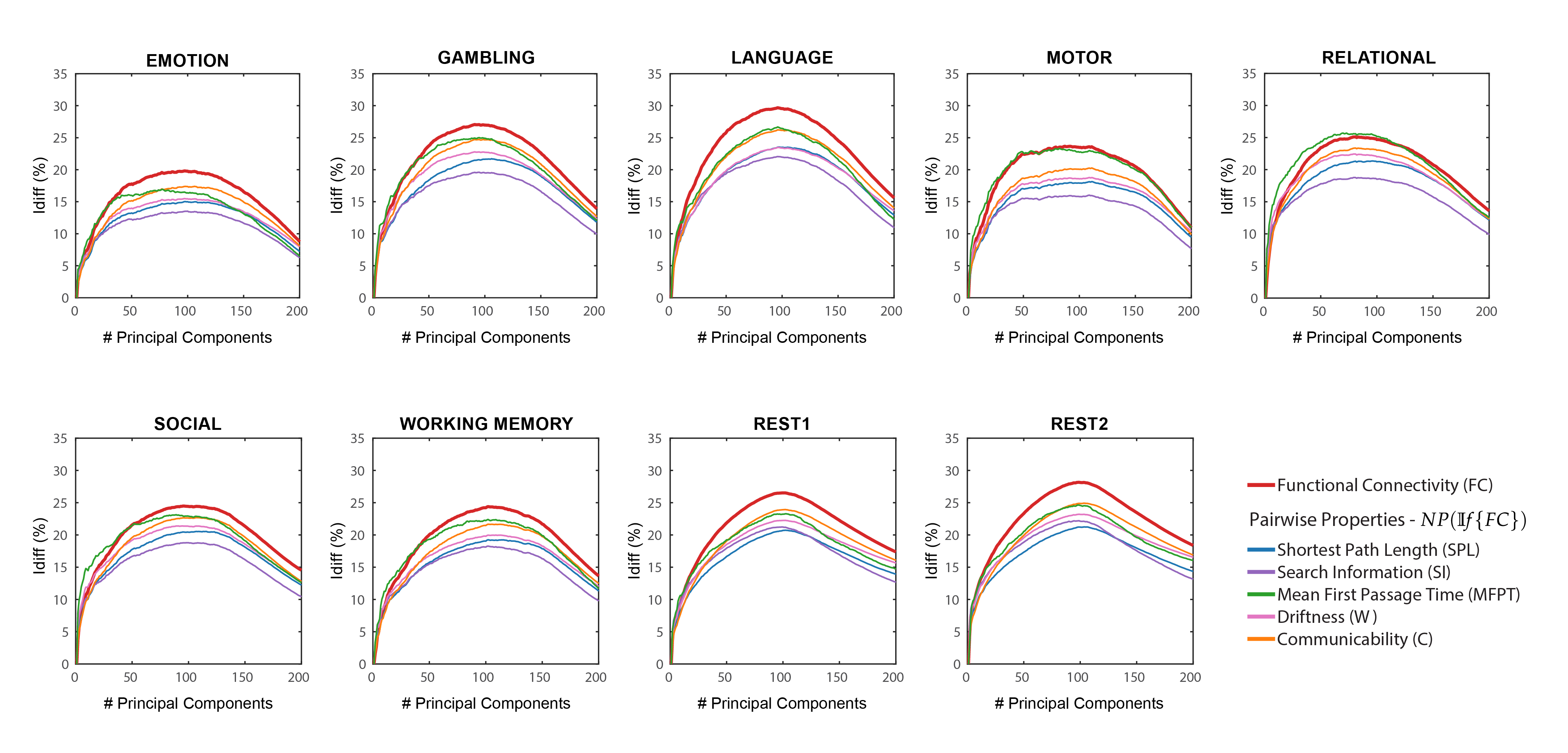}
\caption{$NP(\mathbb{I}\mathit{f}\{FC\})$ \textbf{Differential Identifiability ($\mathit{I_{diff}}$) profiles of pairwise properties} for different fMRI tasks as a function of the number of principal components used for reconstruction. Here, the Identifiability framework was applied on the functional connectomes ($\mathbb{I}\mathit{f}\{FC\}$). Each plot shows, for each fMRI task, the $\mathit{I_{diff}}$ score associated with functional connectivity \textit{(red solid line)} and the $\mathit{I_{diff}}$ scores on network properties derived from the reconstructed functional connectomes, $NP(\mathbb{I}\mathit{f}\{FC\})$ \textit{(see legend)} for different number of components.}
\label{f1}
\end{figure}

For each task, we observed an optimal point of reconstruction where the differential identifiability on the FCs was maximized (see Figure 1 \ref{f1}). This optimal point was always in the neighborhood of half the maximum number of components (which is equal to the number of subjects in the data) and produced $\mathit{I_{diff}}$ values much higher than fully reconstructed data, i.e. using all the components. These results reaffirm those reported by \cite{amico2018quest}. We then assessed $\mathit{I_{diff}}$ on the following node pair network properties: Shortest Path Length (SPL), Search Information (SI), Mean First Passage Time (MFPT), Driftness (W), and Communicability (C). In all cases, there was an optimal regime of number of components that maximized $\mathit{I_{diff}}$ (see Figure 1\ref{f1}). Overall, the $\mathit{I_{diff}}$ score on all the network properties and functional connectomes reach the peak at similar number of principal components, ranging between $80$ and $110$. We can also see that the $\mathit{I_{diff}}$ on functional connectomes is generally higher than those on the network properties for all the tasks and for most of the number of components. One exception is MFPT on Motor task where the $\mathit{I_{diff}}$ scores on FC and MFPT produced very similar results for the entire range of principal components. Another exception is MFPT on Relational task where the peak $\mathit{I_{diff}}$ of $MFPT(\mathbb{I}\mathit{f}\{FC\}$ is greater than that of $\mathbb{I}\mathit{f}\{FC\}$ but the margin of difference is really small ($\approx 0.59$).


\begin{figure}[h]
\includegraphics[width=\textwidth]{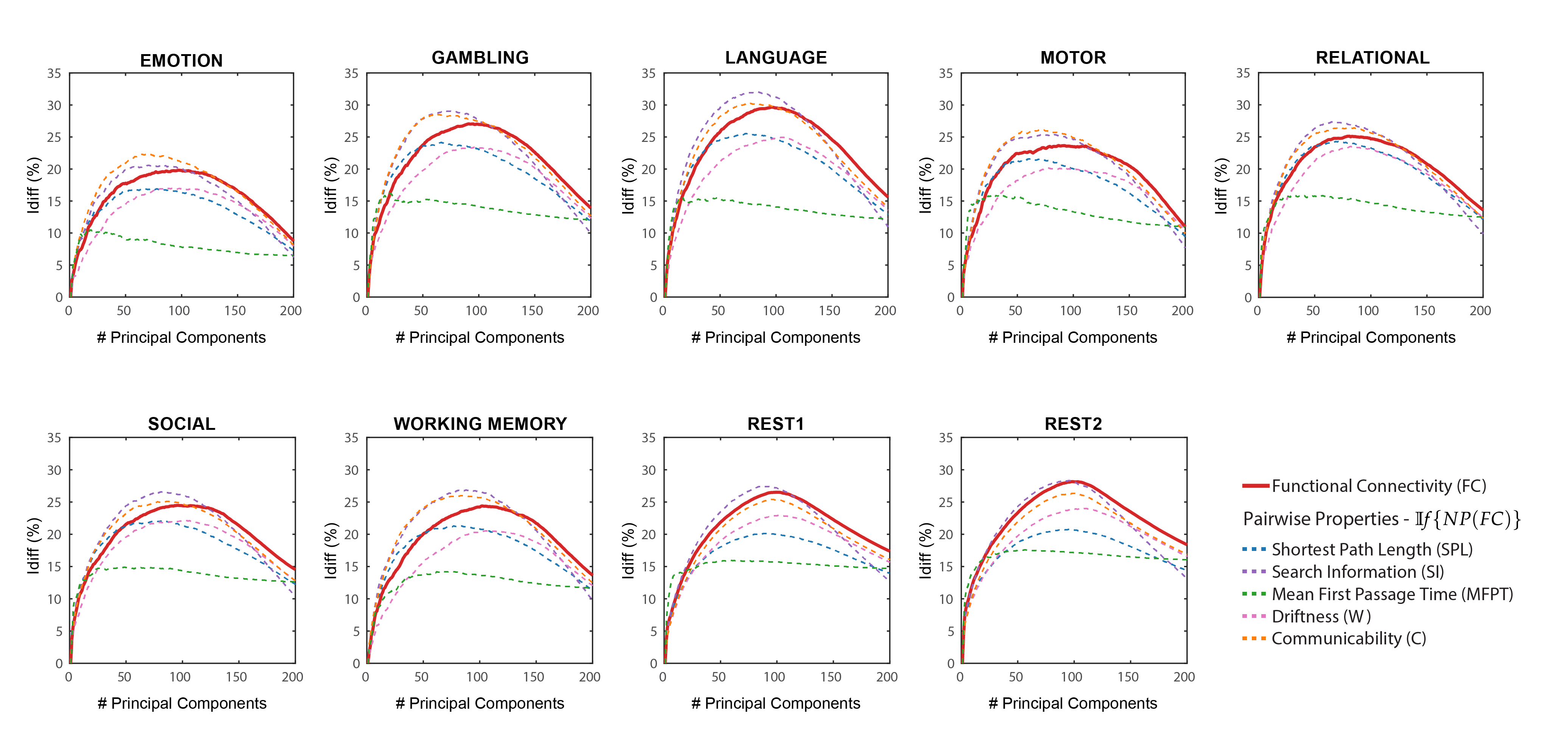}
\caption{$\mathbb{I}\mathit{f}\{NP(FC)\}$ \textbf{Differential Identifiability ($\mathit{I_{diff}}$) profiles of pairwise properties} for different fMRI tasks as a function of the number of principal components used for reconstruction. Here, the Identifiability framework was applied directly on the network properties derived from the original functional connectomes ($\mathbb{I}\mathit{f}\{NP(FC)\}$). Each plot shows, for each fMRI task, the $\mathit{I_{diff}}$ score associated with functional connectivity \textit{(red solid line)} and the $\mathit{I_{diff}}$ scores on reconstructed network properties derived from the original functional connectomes, $\mathbb{I}\mathit{f}\{NP(FC)\}$ \textit{(see legend)} for different number of components.}
\label{f2}
\end{figure}

In $\mathbb{I}\mathit{f}\{NP(FC)\})$ the different network properties (refer Methods) were first derived from the original functional connectomes and subsequently decomposed and reconstructed using the Identifiability framework. $\mathit{I_{diff}}$ scores were computed on these reconstructed network properties for different number of components and compared with those computed from the reconstructed FCs. (see Figure 2\ref{f2})

\begin{figure}[h]
\centering
\includegraphics[width=\textwidth]{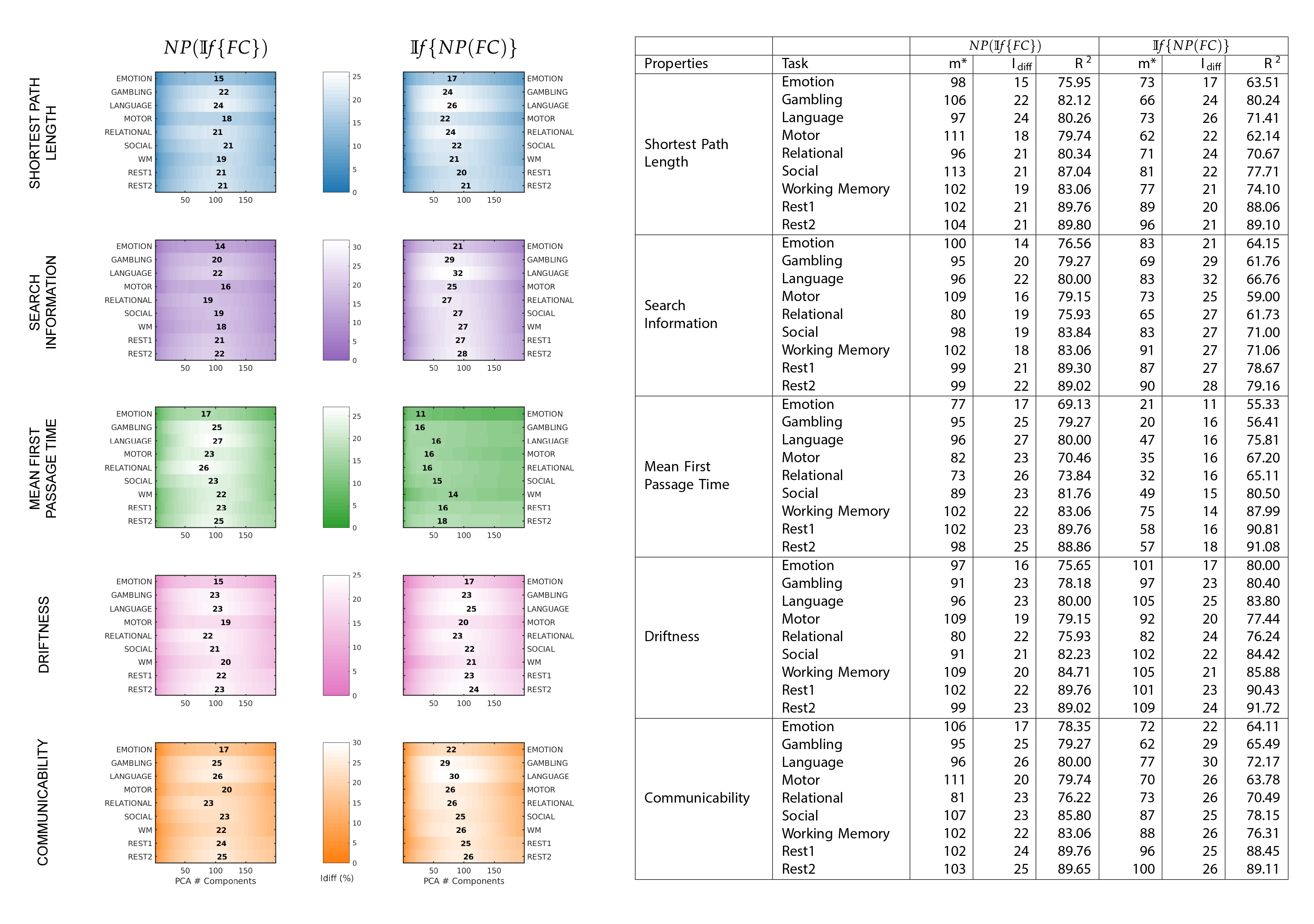}
\caption{A summary of maximum $\mathit{I_{diff}}$ values, corresponding number of components and explained variance retained for each fMRI task and network property for both $NP(\mathbb{I}\mathit{f}\{FC\})$ and $\mathbb{I}\mathit{f}\{NP(FC)\}$. On the left, each plot shows, for each property and each method - $NP(\mathbb{I}\mathit{f}\{FC\})$ or $\mathbb{I}\mathit{f}\{NP(FC)\}$, the $\mathit{I_{diff}}$ score for all tasks. The number mentioned gives the maximum $\mathit{I_{diff}}$ score for the corresponding task \textit{(y axis)} and the position denotes the number of components \textit{(x axis)}. On the left is the same information summarized as a table. For each method and network property, the table gives the number of components used for optimal reconstruction \textit{$m^*$}, corresponding maximum $\mathit{I_{diff}}$ value and the variance explained at that reconstruction $\mathit{R^2}$}
\label{f3}
\end{figure}

As opposed to results shown in Figure 1\ref{f1} which used $NP(\mathbb{I}\mathit{f}\{FC\})$, network properties have heterogeneous $\mathit{I_{diff}}$ profiles with respect to number of components.
Compared to $\mathit{I_{diff}}$ from $\mathbb{I}\mathit{f}\{FC\}$, Search Information has a higher peak $\mathit{I_{diff}}$ score for all tasks while Communicability has a higher peak $\mathit{I_{diff}}$ score for all tasks except resting state.
We also find that MFPT has a very different $\mathit{I_{diff}}$ profile compared to other network properties. The $\mathit{I_{diff}}$ profiles of MFPT from $\mathbb{I}\mathit{f}\{MFPT(FC)\}$ increases as we add the first few component and saturates or decreases gradually as more components are added (starting around $20$ components for all tasks).
This is unlike other network properties and functional connectome that share similar $\mathit{I_{diff}}$ profiles (see Figure 2\ref{f2}). 
A summary of maximum $\mathit{I_{diff}}$, corresponding number of components used and variance retained for $NP(\mathbb{I}\mathit{f}\{FC\})$, and $\mathbb{I}\mathit{f}\{NP(FC)\}$ can be seen in Figure 3\ref{f3}.

\begin{figure}[h!]
\centering
\includegraphics[width=\textwidth]{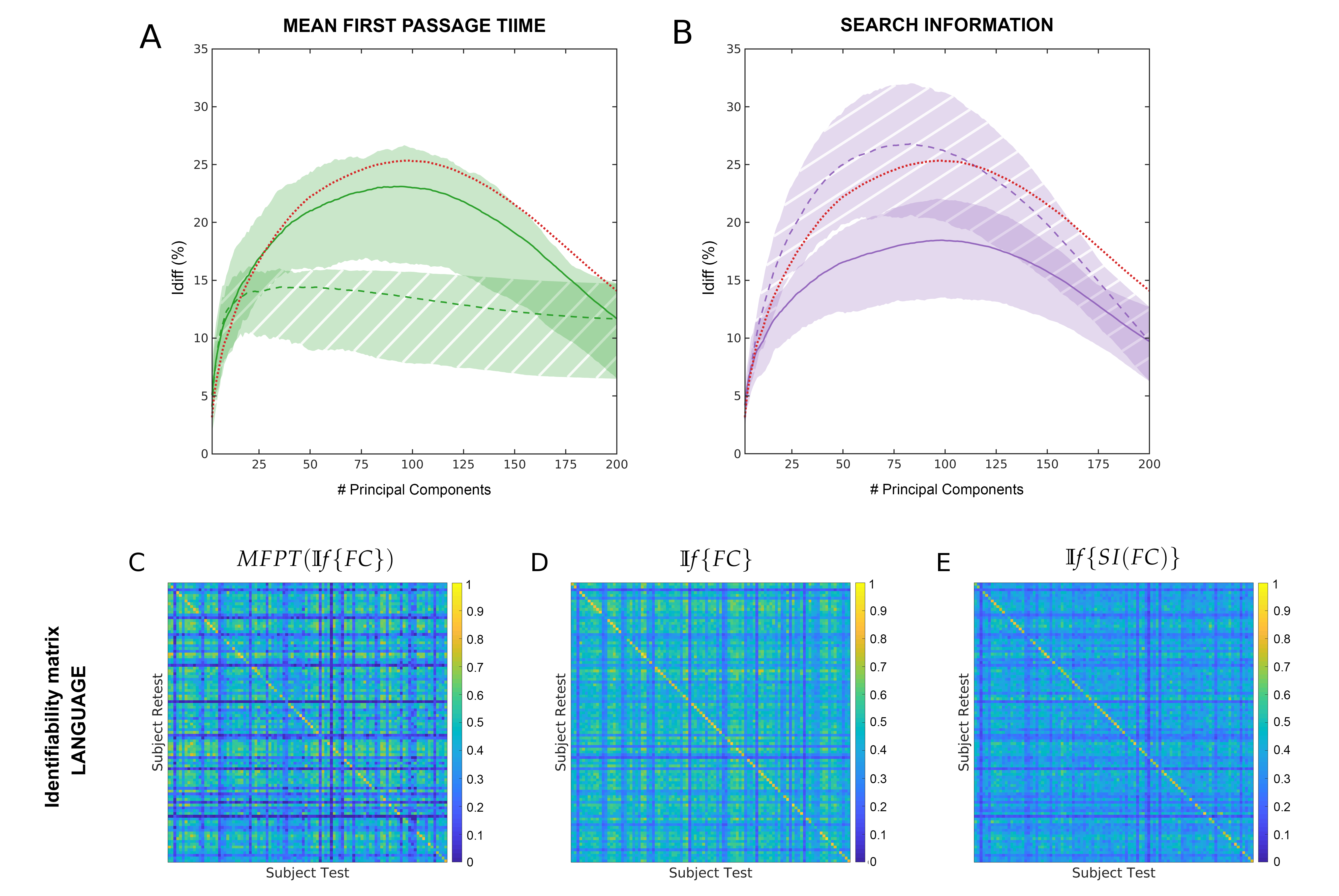}
\caption{(A) Across tasks and rest differential Identifiability ($\mathit{I_{diff}}$) for Mean First Passage Time as a function of the number of principal components used for reconstruction. \textit{Solid line} and \textit{solid shaded area} represent the results for $MFPT(\mathbb{I}\mathit{f}\{FC\})$. \textit{Dashed line} and \textit{hatched area} show results for $\mathbb{I}\mathit{f}\{MFPT(FC)\}$ (B) Across tasks and rest differential Identifiability ($\mathit{I_{diff}}$) for Search Information as a function of the number of principal components used for reconstruction. \textit{Solid line} and \textit{solid shaded area} represent the results for $SI(\mathbb{I}\mathit{f}\{FC\})$. \textit{Dashed line} and \textit{hatched area} show results for $\mathbb{I}\mathit{f}\{SI(FC)\}$ The differential identifiability matrix (as defined in Methods) is shown at optimal reconstruction for Language task for (C) $MFPT(\mathbb{I}\mathit{f}\{FC\})$, (D) $\mathbb{I}\mathit{f}\{FC\}$ and (E) $\mathbb{I}\mathit{f}\{SI(FC)\}$. The diagonal elements in each matrix represent $\mathit{I_{self}}$ and the non-diagonal elements represent $\mathit{I_{others}}$.}
\label{f4}
\end{figure}

The network property with the most different $\mathit{I_{diff}}$ profiles was between $MFPT(\mathbb{I}\mathit{f}\{FC\})$ and $\mathbb{I}\mathit{f}\{MFPT(FC)\}$. Search Information was the only network property that reached higher $\mathit{I_{diff}}$ values for all fMRI tasks for $\mathbb{I}\mathit{f}\{SI(FC)\}$. The difference between Search Information and Mean First Passage time are assessed in detail in Figure 4\ref{f4}. Shaded area highlights the variability of $\mathit{I_{diff}}$ scores across different tasks for $NP(\mathbb{I}\mathit{f}\{FC\})$ (solid area) and $\mathbb{I}\mathit{f}\{NP(FC)\}$ (hatched area). Across all tasks, $\mathit{I_{diff}}$ on $\mathbb{I}\mathit{f}\{SI(FC)\}$ is higher than $SI(\mathbb{I}\mathit{f}\{FC\}$. However, for Mean First Passage time, $\mathit{I_{diff}}$ on $MFPT(\mathbb{I}\mathit{f}\{(FC)\}$ is higher than $(\mathbb{I}\mathit{f}\{MFPT(FC)\}$. When $SI(\mathbb{I}\mathit{f}\{FC\})$ is derived and optimally reconstructed, $\mathit{I_{diff}}$ on Search Information is highest across all tasks. However, under full reconstruction $m=200$ (which is equivalent to using the original functional connectomes), $\mathit{I_{diff}}$ scores are highest for the functional connectome for all fMRI tasks.

\begin{figure}[h!]
\centering
\includegraphics[width=\textwidth]{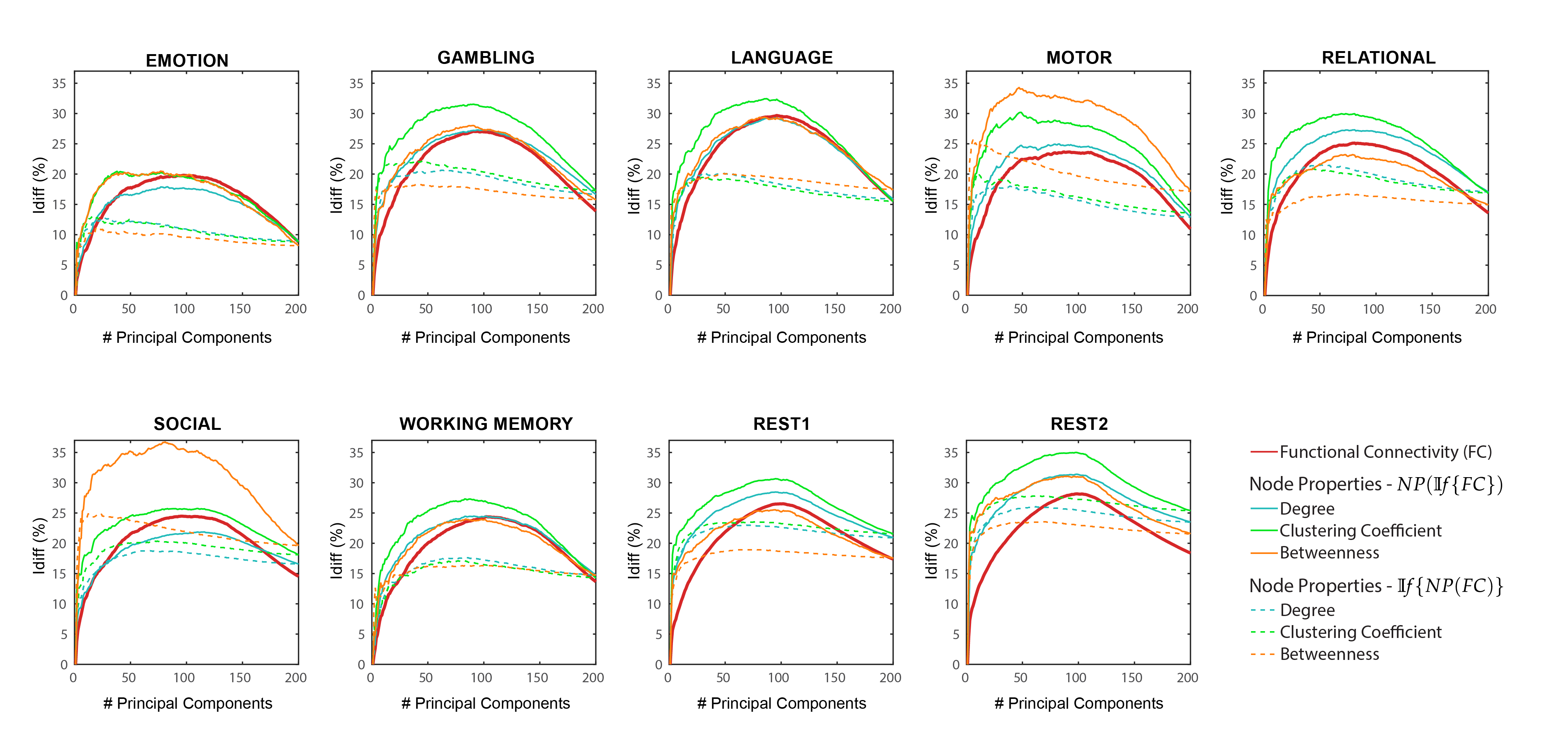}
\caption{$NP(\mathbb{I}\mathit{f}\{FC\})$ and $\mathbb{I}\mathit{f}\{NP(FC)\}$ \textbf{Differential Identifiability ($\mathit{I_{diff}}$) of node properties} for different fMRI tasks as a function of the number of principal components used for reconstruction. Each plot shows, for each task, the $\mathit{I_{diff}}$ score associated with functional connectivity \textit{(red solid line)}, the $\mathit{I_{diff}}$ scores on the network properties derived from the reconstructed functional connectomes $NP(\mathbb{I}\mathit{f}\{FC\})$ \textit{(solid lines, colors - see legend)} and the $\mathit{I_{diff}}$ scores on the reconstructed network properties derived from the original functional connectomes $\mathbb{I}\mathit{f}\{NP(FC)\}$ \textit{(dotted lines, colors - see legend)} for different number of components.}
\label{f5}
\end{figure}

We then assessed how differential identifiability varies based on node properties - Degree, Betweeness Centrality and Clustering Coefficient (Figure 5\ref{f5}). We find that the $\mathit{I_{diff}}$ profiles of $NP(\mathbb{I}\mathit{f}\{FC\}$ are similar to that of $\mathbb{I}\mathit{f}\{FC\}$. These also give a significantly higher optimal $\mathit{I_{diff}}$ score for Gambling, Language, Motor and Working Memory tasks for all node properties. Especially in the case of Language and Motor tasks, Betweeness Centrality gives a significantly higher $\mathit{I_diff}$ of $37$ and $35$ respectively at optimal reconstruction. For $\mathbb{I}\mathit{f}\{NP(FC)\}$, results show lower and flatter $\mathit{I_{diff}}$ profiles for all tasks and a wide range of number of components. $\mathit{I_{diff}}$ profiles using $NP(\mathbb{I}\mathit{f}\{FC\})$ of these node properties are in agreement with all pairwise properties explored so far. In contrast, the $\mathit{I_{diff}}$ profiles using $\mathbb{I}\mathit{f}\{NP(FC)\}$ on these node properties are similar to $\mathbb{I}\mathit{f}\{MFPT(FC)\}$ only.

Intraclass Correlation Coefficient was used to assess the task sensitivity of each pairwise network property for three possible cases - $NP(\mathbb{I}\mathit{f}\{FC\})$ vs $NP(FC)$ (row a), $\mathbb{I}\mathit{f}\{NP(FC)\}$ vs $NP(FC)$ (row b) and $NP(\mathbb{I}\mathit{f}\{FC\})$ vs $\mathbb{I}\mathit{f}\{NP(FC)\}$ (row c). We find that the task sensitivity is higher for all network properties when the Identifiability framework was used (for both $NP(\mathbb{I}\mathit{f}\{FC\})$ and $\mathbb{I}\mathit{f}\{NP(FC)\}$). Between $NP(\mathbb{I}\mathit{f}\{FC\})$ and $\mathbb{I}\mathit{f}\{NP(FC)\}$, there is no one method that improves task sensitivity for all network properties.

\begin{figure}[h!]
\centering
\includegraphics[width=\textwidth]{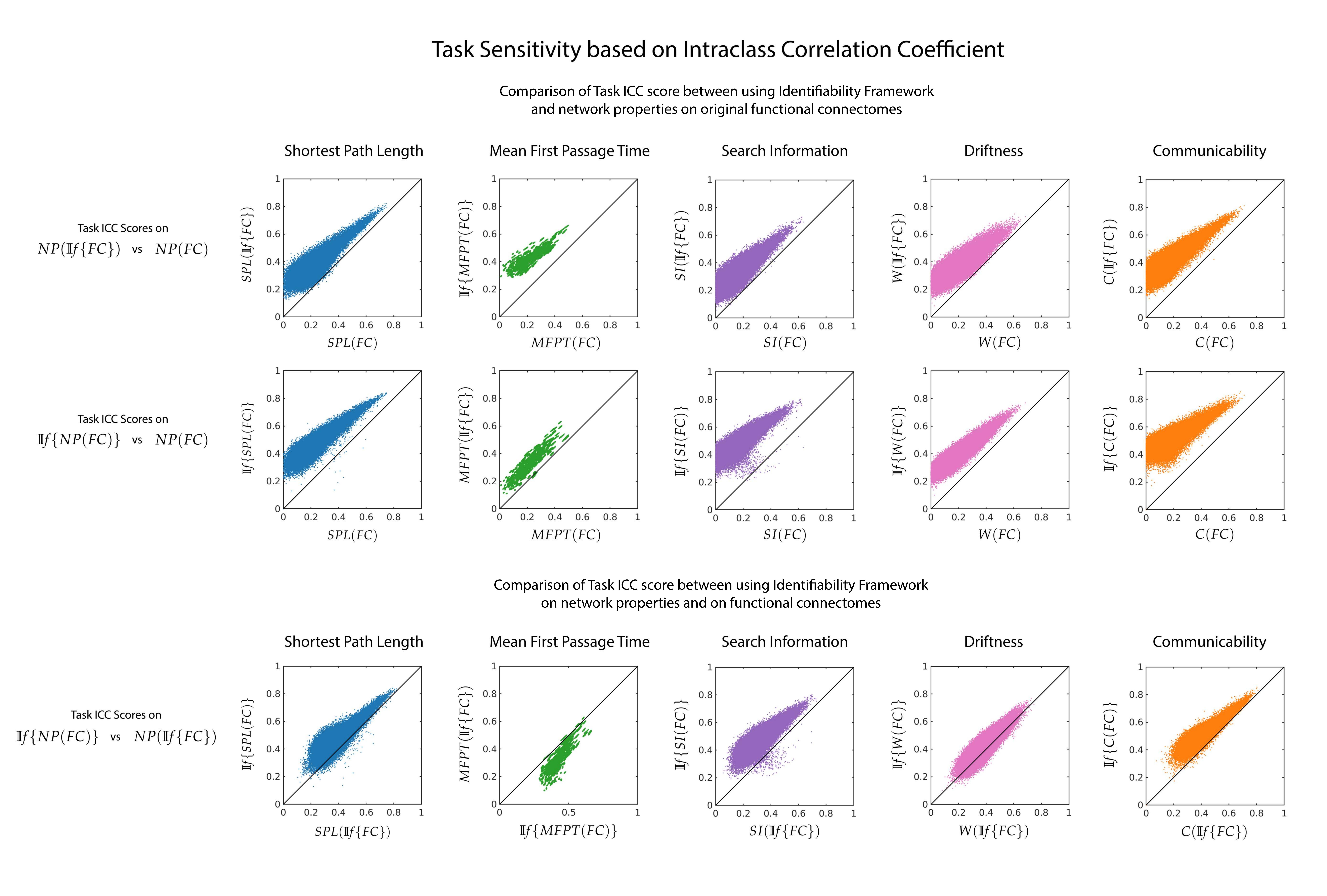}
\caption{Effect of $\mathbb{I}\mathit{f}$ on task sensitivity of network measures. For each pairwise network property, task sensitivity is measured using ICC between - $NP(\mathbb{I}\mathit{f}\{FC\})$ vs $NP(FC)$ (row a), $\mathbb{I}\mathit{f}\{NP(FC)\}$ vs $NP(FC)$ (row b) and $NP(\mathbb{I}\mathit{f}\{FC\})$ vs $\mathbb{I}\mathit{f}\{NP(FC)\}$ (row c). First two rows highlight the fact that the $\mathbb{I}\mathit{f}$ framework uncovers the inherently distinct signature of different tasks through derived network properties. The last row shows that certain network properties would benefit more from application of the $\mathbb{I}\mathit{f}$ framework on the functional connectomes, while others from application directly on the network properties.}
\label{f6}
\end{figure}


\section{Discussion}

Brain connectivity fingerprinting has taken center stage in the neuroscientific community \cite{byrge2019high, miranda2014connectotyping, mars2018connectivity, finn2015functional, satterthwaite2018personalized, gratton2018functional, seitzman2019trait, venkatesh2019comparing}. As we move in this direction, there is a need to improve the reliability and robustness of individual fingerprint in functional connectomes and on common network measures extracted from functional connectomes. The Identifiability Framework ($\mathbb{I}\mathit{f}$) has shown the capacity to uncover subject fingerprint as  measured by $\mathit{I_{diff}}$ score in human functional connectomes, regardless of the fMRI task \cite{amico2018quest}. Improving differential identifiability using the $\mathbb{I}\mathit{f}$ framework on functional connectomes (FCs) has been shown to improve the test-retest reliability of FCs and correlation with fluid intelligence \cite{amico2018quest}. Here, we extend this framework to show that by maximizing individual fingerprints in the functional connectomes, we also maximize individual fingerprint in network properties derived from the connectomes. Furthermore, we found that uncovering individual fingerprinting on network measurements also improves task signature. In addition, we show that in certain network properties, we can uncover an even stronger fingerprint if we apply the framework directly on the network property instead of functional connectomes.

Numerous work has been done to assess the effect of change in parameters of the acquisition process and the preprocessing pipelines on test-retest (TRT) reliability of fMRI data \cite{noble2019decade, shah2016reliability, birn2013effect, noble2017influences}. The impact of different correlation metrics, inclusion or exclusion of edges on functional connectomes, as well as the use of global signal regression, have been explored extensively \cite{liang2012effects, fiecas2013quantifying, schwarz2011negative, wang2011graph, cao2014test, byrge2019high}. Additionally, TRT reliability is also seen to be affected by band pass filtering, scan length, sampling rate, network definition of the weights, and size of voxels for node definition \cite{braun2012test, liang2012effects, liao2013functional}. The fact that the TRT reliability of the fMRI data and the subsequent estimation of functional connectomes is affected by such diverse factors, it is important to explore the reliability of the derived network properties. Even though TRT reliability is not the only parameter to take into account when choosing the optimal strategy for brain network analyses, it surely has to be considered an important factor to help in such an important choice.

Essentially, $\mathbb{I}\mathit{f}$ works as a group-level data-driven (\textit{denoising}) procedure where the components not contributing towards test-retest reliability of FCs are identified and removed. $\mathbb{I}\mathit{f}$ doesn't just improve the overall TRT reliability of a functional connectome but also improves it locally on an edge-level \cite{amico2018quest} which should ensure that both global and local network properties computed using these denoised functional connectomes are more reliable and robust. As shown in Figure 1\ref{f1}, $\mathbb{I}\mathit{f}$ not only maximizes subject fingerprint at the FC level, but also at the network property level, which validated our premise. In addition, this convergent behavior is not present just at the optimal point; the identifiability profile of network properties follows the identifiability profile of the functional connectomes. In essence, we have shown that regardless of whether you are using functional connectomes or the network properties derived from them, using $\mathbb{I}\mathit{f}$ framework on the functional connectomes would be a beneficial first step.


A natural next question was to find if $\mathbb{I}\mathit{f}$ should be applied on functional connectomes and then derive the network properties ($NP(\mathbb{I}\mathit{f}\{FC\})$), or to use it directly on the network properties derived from original functional connectomes ($\mathbb{I}\mathit{f}\{NP(FC)\}$). The two approaches are an attempt to understand different principles of what a fingerprint is in a network derived measurement. $\mathbb{I}\mathit{f}\{NP(FC)\}$ assumes that functional connectomes are "holding" the individual fingerprints and then propagating them to the network measurements. The fact that maximizing fingerprint of functional connectomes also maximizes the fingerprint in derived network measures, suggests that functional connectomes do indeed hold a subject fingerprint which is then transmitted to the derived network properties. On the other hand, we also see that for some network measures (e.g. Search Information), we can uncover a better fingerprint if we apply the framework directly on the network measure. This suggests that specific network measures have a subject fingerprint of their own which gets added on to the functional connectome fingerprint. Hence, if under some circumstances, the goal is to maximize the reliability and the individual variability of a specific network property, one can benefit from applying the $\mathbb{I}\mathit{f}$ framework on the network property itself, rather than on FCs.

Notably, in the $\mathbb{I}\mathit{f}\{SI(FC)\}$ scenario, the most different $\mathit{I_{diff}}$ profiles were found between MFPT and Search Information (Figure 4\ref{f4}). Search Information consistently provides a better fingerprint across all tasks than functional connectome. MFPT, however, can neither improve nor match the fingerprint of functional connectomes. Also, it can not retain the fingerprint that is otherwise present is the functional connectomes and is then propgated to MFPT using $\mathbb{I}\mathit{f}\{MFPT(FC)\}$. Hence, while some properties (i.e., Search Information) can derive higher identifiability than functional connectomes, properties like MFPT need to be computed on optimally reconstructed functional connectomes to uncover subject identifiability on it.

These findings show that brain fingerprinting can be improved by adding multivariate information to ``bivariate'' measurements such as pairwise correlation used to estimate FCs. Specifically, individual fingerprint peaks on network measurements (e.g. Search Information) that are more multivariate and requires more information on the global topology of the functional network. However, if the information is heavily driven by degree properties (e.g. MFPT), then there is no improvement on the individual fingerprint (Figure 4\ref{f4}). This is strongly corroborated by the $\mathit{I_{diff}}$ profiles of several node properties under the $\mathbb{I}\mathit{f}\{NP(FC)\}$ scenario. These profiles are very similar to that of MFPT, a network property which has a strong negative correlation with the degree of the target node. Although $\mathbb{I}\mathit{f}\{NP(FC)\}$ of these node properties have $\mathit{I_{diff}}$ profiles similar to $\mathbb{I}\mathit{f}\{MFPT(FC)\}$, the maximum $\mathit{I_{diff}}$ on these node properties are, for some tasks, significantly higher than $\mathbb{I}\mathit{f}\{FC\}$. Betweeness Centrality, for example, has a higher subject identifiability for Social and Motor tasks.

It was interesting to observe that under the $\mathbb{I}\mathit{f}\{NP(FC)\}$ scenario, Betweenness Centrality maximizes differential identifiability using just the first two components for Social and Motor tasks and that it was higher than the identifiability of the functional connectomes for any number of components. Since Betweenness Centrality can be used to identify integrative communication hubs in FCs \cite{sporns2013network}, it can be argued that social and motor tasks display a ``hub functional fingerprint", which can be captured by the first two principal components. 


A complementary assessment to the identification of subject fingerprints is to assess the ability to identify the different tasks used in this study. To do so, we used intraclass correlation coefficient on the derived network properties. The $\mathbb{I}\mathit{f}$ framework improved task sensitivity on the network properties (see Figure 6\ref{f6}). Regardless of using the framework on the original functional connectomes or on the network properties themselves, a higher task sensitivity is obtained using one of the process depending on the network property. In both cases, the task reliability of the network properties has improved. The different tasks in the HCP dataset aim to assess different cognitive processes. Hence, the corresponding connectomes and the network properties derived from them should, at least to some extent, be task specific. We have shown that using the $\mathbb{I}\mathit{f}$ framework uncovers task-related fingerprints where unique cognitive processes result in differential network properties.



To summarize, differential identifiability was found to be always higher on functional connectomes than on any network properties when the Identifiability framework ($\mathbb{I}\mathit{f}$) is not used. When $\mathbb{I}\mathit{f}$ improved identifiability on functional connectomes, the identifiability on the network properties also increased. The framework also improved the subject fingerprints of the network properties. Not only do they improve at the optimal point, but the differential identifiability follows the same profile on network properties as it does on functional connectomes. We also find that applying the identifiability framework on the network properties instead of functional connectomes gives higher differential identifiability for some network properties. At optimal reconstruction, we find that Search Information has higher differential identifiability than functional connectomes across all tasks when the identifiability framework is applied on search information. This shows that there are network properties that can uncover better identifiability with framework than the functional connectomes themselves. Finally, we found that using the identifiability framework (either on functional connectomes or network property) improves task sensitivity in all network properties.


Only the unrelated subjects of the Human Connectome project and the cortical parcellation proposed by \cite{glasser2013minimal} are used in this work. Other explorations with other atlases, parcellations and/or other estimators of functional coupling (other than Pearson's correlation coefficient) would expand on the implications of our work. We have also limited to commonly used five pairwise and three node network properties. Delving into other network properties can strengthen this framework further and provide additional insights in understanding the associations between brain fingerprints, functional connectivity, and network derived properties.

This study can be extended to clinical applications to understand diseases that target specific functions of the human brain. Pathology whose signature cannot be mapped on the functional connectome itself but can be assessed using different network properties. \cite{bassett2009human, fornito2015connectomic, fornito2015connectomics} In this case, to retain individual differences and to be able to differentiate healthy population from clinical ones, we need this study to understand the advantages of using the Identifiability framework on the functional connectome or network property. Finally, studying the effect of the framework on the structural connectome is another natural extension of this work.

\bibliographystyle{unsrt}  
\bibliography{references}  

\begin{thebibliography}{10}

\bibitem{sporns2010networks}
Olaf Sporns.
\newblock {\em Networks of the Brain}.
\newblock MIT press, 2010.

\bibitem{fornito2016fundamentals}
Alex Fornito, Andrew Zalesky, and Edward Bullmore.
\newblock {\em Fundamentals of brain network analysis}.
\newblock Academic Press, 2016.

\bibitem{sporns2018graph}
Olaf Sporns.
\newblock Graph theory methods: applications in brain networks.
\newblock {\em Dialogues in clinical neuroscience}, 20(2):111, 2018.

\bibitem{sporns2016modular}
Olaf Sporns and Richard~F Betzel.
\newblock Modular brain networks.
\newblock {\em Annual review of psychology}, 67:613--640, 2016.

\bibitem{cohen2016segregation}
Jessica~R Cohen and Mark D'Esposito.
\newblock The segregation and integration of distinct brain networks and their
  relationship to cognition.
\newblock {\em Journal of Neuroscience}, 36(48):12083--12094, 2016.

\bibitem{deco2015rethinking}
Gustavo Deco, Giulio Tononi, Melanie Boly, and Morten~L Kringelbach.
\newblock Rethinking segregation and integration: contributions of whole-brain
  modelling.
\newblock {\em Nature Reviews Neuroscience}, 16(7):430--439, 2015.

\bibitem{fukushima2018structure}
Makoto Fukushima, Richard~F Betzel, Ye~He, Martijn~P van~den Heuvel, Xi-Nian
  Zuo, and Olaf Sporns.
\newblock Structure--function relationships during segregated and integrated
  network states of human brain functional connectivity.
\newblock {\em Brain Structure and Function}, 223(3):1091--1106, 2018.

\bibitem{sporns2013network}
Olaf Sporns.
\newblock Network attributes for segregation and integration in the human
  brain.
\newblock {\em Current opinion in neurobiology}, 23(2):162--171, 2013.

\bibitem{costa2011communication}
Luciano da~Fontoura Costa, Jo{\~a}o~LB Batista, and Giorgio~A Ascoli.
\newblock Communication structure of cortical networks.
\newblock {\em Frontiers in computational neuroscience}, 5:6, 2011.

\bibitem{estrada2008communicability}
Ernesto Estrada and Naomichi Hatano.
\newblock Communicability in complex networks.
\newblock {\em Physical Review E}, 77(3):036111, 2008.

\bibitem{petrella2011use}
Jeffrey~R Petrella.
\newblock Use of graph theory to evaluate brain networks: a clinical tool for a
  small world?, 2011.

\bibitem{avena2018communication}
Andrea Avena-Koenigsberger, Bratislav Misic, and Olaf Sporns.
\newblock Communication dynamics in complex brain networks.
\newblock {\em Nature Reviews Neuroscience}, 19(1):17, 2018.

\bibitem{zalesky2010network}
Andrew Zalesky, Alex Fornito, and Edward~T Bullmore.
\newblock Network-based statistic: identifying differences in brain networks.
\newblock {\em Neuroimage}, 53(4):1197--1207, 2010.

\bibitem{davison2015brain}
Elizabeth~N Davison, Kimberly~J Schlesinger, Danielle~S Bassett, Mary-Ellen
  Lynall, Michael~B Miller, Scott~T Grafton, and Jean~M Carlson.
\newblock Brain network adaptability across task states.
\newblock {\em PLoS computational biology}, 11(1):e1004029, 2015.

\bibitem{bola2015dynamic}
Micha{\l} Bola and Bernhard~A Sabel.
\newblock Dynamic reorganization of brain functional networks during cognition.
\newblock {\em Neuroimage}, 114:398--413, 2015.

\bibitem{alavash2015persistency}
Mohsen Alavash, Claus~C Hilgetag, Christiane~M Thiel, and Carsten Gie{\ss}ing.
\newblock Persistency and flexibility of complex brain networks underlie
  dual-task interference.
\newblock {\em Human brain mapping}, 36(9):3542--3562, 2015.

\bibitem{mattar2016flexible}
Marcelo~G Mattar, Richard~F Betzel, and Danielle~S Bassett.
\newblock The flexible brain.
\newblock {\em Brain}, 139(8):2110--2112, 2016.

\bibitem{fornito2015connectomics}
Alex Fornito, Andrew Zalesky, and Michael Breakspear.
\newblock The connectomics of brain disorders.
\newblock {\em Nature Reviews Neuroscience}, 16(3):159--172, 2015.

\bibitem{castellanos2013clinical}
F~Xavier Castellanos, Adriana Di~Martino, R~Cameron Craddock, Ashesh~D Mehta,
  and Michael~P Milham.
\newblock Clinical applications of the functional connectome.
\newblock {\em Neuroimage}, 80:527--540, 2013.

\bibitem{crossley2014hubs}
Nicolas~A Crossley, Andrea Mechelli, Jessica Scott, Francesco Carletti, Peter~T
  Fox, Philip McGuire, and Edward~T Bullmore.
\newblock The hubs of the human connectome are generally implicated in the
  anatomy of brain disorders.
\newblock {\em Brain}, 137(8):2382--2395, 2014.

\bibitem{seitzman2019trait}
Benjamin~A Seitzman, Caterina Gratton, Timothy~O Laumann, Evan~M Gordon,
  Babatunde Adeyemo, Ally Dworetsky, Brian~T Kraus, Adrian~W Gilmore, Jeffrey~J
  Berg, Mario Ortega, et~al.
\newblock Trait-like variants in human functional brain networks.
\newblock {\em Proceedings of the National Academy of Sciences},
  116(45):22851--22861, 2019.

\bibitem{van2013wu}
David~C Van~Essen, Stephen~M Smith, Deanna~M Barch, Timothy~EJ Behrens, Essa
  Yacoub, Kamil Ugurbil, Wu-Minn~HCP Consortium, et~al.
\newblock The wu-minn human connectome project: an overview.
\newblock {\em Neuroimage}, 80:62--79, 2013.

\bibitem{biswal2010toward}
Bharat~B Biswal, Maarten Mennes, Xi-Nian Zuo, Suril Gohel, Clare Kelly, Steve~M
  Smith, Christian~F Beckmann, Jonathan~S Adelstein, Randy~L Buckner, Stan
  Colcombe, et~al.
\newblock Toward discovery science of human brain function.
\newblock {\em Proceedings of the National Academy of Sciences},
  107(10):4734--4739, 2010.

\bibitem{mars2018connectivity}
Rogier~B Mars, Richard~E Passingham, and Saad Jbabdi.
\newblock Connectivity fingerprints: from areal descriptions to abstract
  spaces.
\newblock {\em Trends in cognitive sciences}, 22(11):1026--1037, 2018.

\bibitem{satterthwaite2018personalized}
Theodore~D Satterthwaite, Cedric~H Xia, and Danielle~S Bassett.
\newblock Personalized neuroscience: Common and individual-specific features in
  functional brain networks.
\newblock {\em Neuron}, 98(2):243--245, 2018.

\bibitem{gratton2018functional}
Caterina Gratton, Timothy~O Laumann, Ashley~N Nielsen, Deanna~J Greene, Evan~M
  Gordon, Adrian~W Gilmore, Steven~M Nelson, Rebecca~S Coalson, Abraham~Z
  Snyder, Bradley~L Schlaggar, et~al.
\newblock Functional brain networks are dominated by stable group and
  individual factors, not cognitive or daily variation.
\newblock {\em Neuron}, 98(2):439--452, 2018.

\bibitem{venkatesh2019comparing}
Manasij Venkatesh, Joseph Jaja, et~al.
\newblock Comparing functional connectivity matrices: A geometry-aware approach
  applied to participant identification.
\newblock {\em bioRxiv}, page 687830, 2019.

\bibitem{finn2015functional}
Emily~S Finn, Xilin Shen, Dustin Scheinost, Monica~D Rosenberg, Jessica Huang,
  Marvin~M Chun, Xenophon Papademetris, and R~Todd Constable.
\newblock Functional connectome fingerprinting: identifying individuals using
  patterns of brain connectivity.
\newblock {\em Nature neuroscience}, 18(11):1664, 2015.

\bibitem{pallares2018extracting}
Vicente Pallar{\'e}s, Andrea Insabato, Ana Sanju{\'a}n, Simone K{\"u}hn, Dante
  Mantini, Gustavo Deco, and Matthieu Gilson.
\newblock Extracting orthogonal subject-and condition-specific signatures from
  fmri data using whole-brain effective connectivity.
\newblock {\em Neuroimage}, 178:238--254, 2018.

\bibitem{xie2018whole}
Hua Xie, Vince~D Calhoun, Javier Gonzalez-Castillo, Eswar Damaraju, Robyn
  Miller, Peter~A Bandettini, and Sunanda Mitra.
\newblock Whole-brain connectivity dynamics reflect both task-specific and
  individual-specific modulation: A multitask study.
\newblock {\em Neuroimage}, 180:495--504, 2018.

\bibitem{rosenberg2019functional}
Monica~D Rosenberg, Dustin Scheinost, Abigail~S Greene, Emily~W Avery,
  Young~Hye Kwon, Emily~S Finn, Ramachandran Ramani, Maolin Qiu, R~Todd
  Constable, and Marvin~M Chun.
\newblock Functional connectivity predicts changes in attention over minutes,
  days, and months.
\newblock {\em bioRxiv}, page 700476, 2019.

\bibitem{amico2018quest}
Enrico Amico and Joaqu{\'\i}n Go{\~n}i.
\newblock The quest for identifiability in human functional connectomes.
\newblock {\em Scientific reports}, 8(1):8254, 2018.

\bibitem{bari2019uncovering}
Sumra Bari, Enrico Amico, Nicole Vike, Thomas~M Talavage, and Joaqu{\'\i}n
  Go{\~n}i.
\newblock Uncovering multi-site identifiability based on resting-state
  functional connectomes.
\newblock {\em NeuroImage}, 2019.

\bibitem{svaldi2019optimizing}
Diana~O Svaldi, Joaqu{\'\i}n Go{\~n}i, Kausar Abbas, Enrico Amico, David~G
  Clark, Charanya Muralidharan, Mario Dzemidzic, John~D West, Shannon~L
  Risacher, Andrew~J Saykin, et~al.
\newblock Optimizing differential identifiability improves connectome
  predictive modeling of cognitive deficits in alzheimer$\backslash$'s disease.
\newblock {\em arXiv preprint arXiv:1908.06197}, 2019.

\bibitem{svaldi2018towards}
Diana~O Svaldi, Joaqu{\'\i}n Go{\~n}i, Apoorva~Bharthur Sanjay, Enrico Amico,
  Shannon~L Risacher, John~D West, Mario Dzemidzic, Andrew Saykin, and Liana
  Apostolova.
\newblock Towards subject and diagnostic identifiability in the alzheimer’s
  disease spectrum based on functional connectomes.
\newblock In {\em Graphs in Biomedical Image Analysis and Integrating Medical
  Imaging and Non-Imaging Modalities}, pages 74--82. Springer, 2019.

\bibitem{scheinost2019ten}
Dustin Scheinost, Stephanie Noble, Corey Horien, Abigail~S Greene, Evelyn~MR
  Lake, Mehraveh Salehi, Siyuan Gao, Xilin Shen, David O'Connor, Daniel~S
  Barron, et~al.
\newblock Ten simple rules for predictive modeling of individual differences in
  neuroimaging.
\newblock {\em NeuroImage}, 2019.

\bibitem{shen2017using}
Xilin Shen, Emily~S Finn, Dustin Scheinost, Monica~D Rosenberg, Marvin~M Chun,
  Xenophon Papademetris, and R~Todd Constable.
\newblock Using connectome-based predictive modeling to predict individual
  behavior from brain connectivity.
\newblock {\em nature protocols}, 12(3):506, 2017.

\bibitem{glasser2013minimal}
Matthew~F Glasser, Stamatios~N Sotiropoulos, J~Anthony Wilson, Timothy~S
  Coalson, Bruce Fischl, Jesper~L Andersson, Junqian Xu, Saad Jbabdi, Matthew
  Webster, Jonathan~R Polimeni, et~al.
\newblock The minimal preprocessing pipelines for the human connectome project.
\newblock {\em Neuroimage}, 80:105--124, 2013.

\bibitem{amico2018mapping}
Enrico Amico and Joaqu{\'\i}n Go{\~n}i.
\newblock Mapping hybrid functional-structural connectivity traits in the human
  connectome.
\newblock {\em Network Neuroscience}, 2(3):306--322, 2018.

\bibitem{smith2013resting}
Stephen~M Smith, Christian~F Beckmann, Jesper Andersson, Edward~J Auerbach,
  Janine Bijsterbosch, Gwena{\"e}lle Douaud, Eugene Duff, David~A Feinberg,
  Ludovica Griffanti, Michael~P Harms, et~al.
\newblock Resting-state fmri in the human connectome project.
\newblock {\em Neuroimage}, 80:144--168, 2013.

\bibitem{amico2019centralized}
Enrico Amico, Alex Arenas, and Joaqu{\'\i}n Go{\~n}i.
\newblock Centralized and distributed cognitive task processing in the human
  connectome.
\newblock {\em Network Neuroscience}, 3(2):455--474, 2019.

\bibitem{rubinov2010complex}
Mikail Rubinov and Olaf Sporns.
\newblock Complex network measures of brain connectivity: uses and
  interpretations.
\newblock {\em Neuroimage}, 52(3):1059--1069, 2010.

\bibitem{rosvall2005networks}
Martin Rosvall, Ala Trusina, Petter Minnhagen, and Kim Sneppen.
\newblock Networks and cities: An information perspective.
\newblock {\em Physical Review Letters}, 94(2):028701, 2005.

\bibitem{goni2014resting}
Joaqu{\'\i}n Go{\~n}i, Martijn~P van~den Heuvel, Andrea Avena-Koenigsberger,
  Nieves~Velez de~Mendizabal, Richard~F Betzel, Alessandra Griffa, Patric
  Hagmann, Bernat Corominas-Murtra, Jean-Philippe Thiran, and Olaf Sporns.
\newblock Resting-brain functional connectivity predicted by analytic measures
  of network communication.
\newblock {\em Proceedings of the National Academy of Sciences},
  111(2):833--838, 2014.

\bibitem{kemeny1976markov}
John~G Kemeny and J~Laurie Snell.
\newblock {\em Markov Chains}.
\newblock Springer-Verlag, New York, 1976.

\bibitem{crofts2009weighted}
Jonathan~J Crofts and Desmond~J Higham.
\newblock A weighted communicability measure applied to complex brain networks.
\newblock {\em Journal of the Royal Society Interface}, 6(33):411--414, 2009.

\bibitem{mcgraw1996forming}
Kenneth~O McGraw and Seok~P Wong.
\newblock Forming inferences about some intraclass correlation coefficients.
\newblock {\em Psychological methods}, 1(1):30, 1996.

\bibitem{bartko1966intraclass}
John~J Bartko.
\newblock The intraclass correlation coefficient as a measure of reliability.
\newblock {\em Psychological reports}, 19(1):3--11, 1966.

\bibitem{shrout1979intraclass}
Patrick~E Shrout and Joseph~L Fleiss.
\newblock Intraclass correlations: uses in assessing rater reliability.
\newblock {\em Psychological bulletin}, 86(2):420, 1979.

\bibitem{glasser2016multi}
Matthew~F Glasser, Timothy~S Coalson, Emma~C Robinson, Carl~D Hacker, John
  Harwell, Essa Yacoub, Kamil Ugurbil, Jesper Andersson, Christian~F Beckmann,
  Mark Jenkinson, et~al.
\newblock A multi-modal parcellation of human cerebral cortex.
\newblock {\em Nature}, 536(7615):171, 2016.

\bibitem{byrge2019high}
Lisa Byrge and Daniel~P Kennedy.
\newblock High-accuracy individual identification using a “thin slice” of
  the functional connectome.
\newblock {\em Network Neuroscience}, 3(2):363--383, 2019.

\bibitem{miranda2014connectotyping}
Oscar Miranda-Dominguez, Brian~D Mills, Samuel~D Carpenter, Kathleen~A Grant,
  Christopher~D Kroenke, Joel~T Nigg, and Damien~A Fair.
\newblock Connectotyping: model based fingerprinting of the functional
  connectome.
\newblock {\em PloS one}, 9(11):e111048, 2014.

\bibitem{noble2019decade}
Stephanie Noble, Dustin Scheinost, and R~Todd Constable.
\newblock A decade of test-retest reliability of functional connectivity: A
  systematic review and meta-analysis.
\newblock {\em Neuroimage}, 203:116157, 2019.

\bibitem{shah2016reliability}
Lubdha~M Shah, Justin~A Cramer, Michael~A Ferguson, Rasmus~M Birn, and
  Jeffrey~S Anderson.
\newblock Reliability and reproducibility of individual differences in
  functional connectivity acquired during task and resting state.
\newblock {\em Brain and behavior}, 6(5):e00456, 2016.

\bibitem{birn2013effect}
Rasmus~M Birn, Erin~K Molloy, R{\'e}mi Patriat, Taurean Parker, Timothy~B
  Meier, Gregory~R Kirk, Veena~A Nair, M~Elizabeth Meyerand, and Vivek
  Prabhakaran.
\newblock The effect of scan length on the reliability of resting-state fmri
  connectivity estimates.
\newblock {\em Neuroimage}, 83:550--558, 2013.

\bibitem{noble2017influences}
Stephanie Noble, Marisa~N Spann, Fuyuze Tokoglu, Xilin Shen, R~Todd Constable,
  and Dustin Scheinost.
\newblock Influences on the test--retest reliability of functional connectivity
  mri and its relationship with behavioral utility.
\newblock {\em Cerebral Cortex}, 27(11):5415--5429, 2017.

\bibitem{liang2012effects}
Xia Liang, Jinhui Wang, Chaogan Yan, Ni~Shu, Ke~Xu, Gaolang Gong, and Yong He.
\newblock Effects of different correlation metrics and preprocessing factors on
  small-world brain functional networks: a resting-state functional mri study.
\newblock {\em PloS one}, 7(3):e32766, 2012.

\bibitem{fiecas2013quantifying}
Mark Fiecas, Hernando Ombao, Dan Van~Lunen, Richard Baumgartner, Alexandre
  Coimbra, and Dai Feng.
\newblock Quantifying temporal correlations: a test--retest evaluation of
  functional connectivity in resting-state fmri.
\newblock {\em NeuroImage}, 65:231--241, 2013.

\bibitem{schwarz2011negative}
Adam~J Schwarz and John McGonigle.
\newblock Negative edges and soft thresholding in complex network analysis of
  resting state functional connectivity data.
\newblock {\em Neuroimage}, 55(3):1132--1146, 2011.

\bibitem{wang2011graph}
Jin-Hui Wang, Xi-Nian Zuo, Suril Gohel, Michael~P Milham, Bharat~B Biswal, and
  Yong He.
\newblock Graph theoretical analysis of functional brain networks: test-retest
  evaluation on short-and long-term resting-state functional mri data.
\newblock {\em PloS one}, 6(7):e21976, 2011.

\bibitem{cao2014test}
Hengyi Cao, Michael~M Plichta, Axel Sch{\"a}fer, Leila Haddad, Oliver Grimm,
  Michael Schneider, Christine Esslinger, Peter Kirsch, Andreas
  Meyer-Lindenberg, and Heike Tost.
\newblock Test--retest reliability of fmri-based graph theoretical properties
  during working memory, emotion processing, and resting state.
\newblock {\em Neuroimage}, 84:888--900, 2014.

\bibitem{braun2012test}
Urs Braun, Michael~M Plichta, Christine Esslinger, Carina Sauer, Leila Haddad,
  Oliver Grimm, Daniela Mier, Sebastian Mohnke, Andreas Heinz, Susanne Erk,
  et~al.
\newblock Test--retest reliability of resting-state connectivity network
  characteristics using fmri and graph theoretical measures.
\newblock {\em Neuroimage}, 59(2):1404--1412, 2012.

\bibitem{liao2013functional}
Xu-Hong Liao, Ming-Rui Xia, Ting Xu, Zheng-Jia Dai, Xiao-Yan Cao, Hai-Jing Niu,
  Xi-Nian Zuo, Yu-Feng Zang, and Yong He.
\newblock Functional brain hubs and their test--retest reliability: a multiband
  resting-state functional mri study.
\newblock {\em Neuroimage}, 83:969--982, 2013.

\bibitem{bassett2009human}
Danielle~S Bassett and Edward~T Bullmore.
\newblock Human brain networks in health and disease.
\newblock {\em Current opinion in neurology}, 22(4):340, 2009.

\bibitem{fornito2015connectomic}
Alex Fornito and Edward~T Bullmore.
\newblock Connectomics: a new paradigm for understanding brain disease.
\newblock {\em European Neuropsychopharmacology}, 25(5):733--748, 2015.

\end{thebibliography}


\end{document}